# Accelerating Latency-Critical Applications with AI-Powered Semi-Automatic Fine-Grained Parallelization on SMT Processors

Denis Los, Igor Petushkov

*Abstract*—Latency-critical applications tend to show low utilization of functional units due to frequent cache misses and mispredictions during speculative execution in high-performance superscalar processors. However, due to significant impact on single-thread performance, Simultaneous Multithreading (SMT) technology is rarely used with heavy threads of latency-critical applications. In this paper, we explore utilization of SMT technology to support fine-grained parallelization of latency-critical applications. Following the advancements in the development of Large Language Models (LLMs), we introduce Aira, an AI-powered Parallelization Adviser. To implement Aira, we extend AI Coding Agent in Cursor IDE with additional tools connected through Model Context Protocol, enabling end-to-end AI Agent for parallelization. Additional connected tools enable LLM-guided hotspot detection, collection of dynamic dependencies with Dynamic Binary Instrumentation, SMT-aware performance simulation to estimate performance gains. We apply Aira with Relic parallel framework for fine-grained task parallelism on SMT cores to parallelize latency-critical benchmarks representing real-world applications used in industry. We show 17% geomean performance gain from parallelization of latency-critical benchmarks using Aira with Relic framework.

## I. INTRODUCTION

Latency-critical applications require strict adherence to timing constraints for response times. Failure to meet these constraints may significantly degrade the user experience, cause system failure, or pose a serious safety threat. Latency-critical applications span over a wide range of domains, such as finance, cloud computing, healthcare, robotics, autonomous systems, aerospace, online gaming, and telecommunications. Modern latency-critical services process more than a billion requests each day.

To meet new demands, major technology companies allocate considerable time and resources to enhance the performance, power efficiency, and effectiveness of latency-critical applications. Modern high-performance processors utilize Out-of-Order execution and superscalar architecture to exploit Instruction-Level Parallelism (ILP) and boost performance. However, even with all the resources invested in performance optimization, latency-critical applications due to frequent cache misses or branch mispredictions could have a low ILP, hence, underutilizing available functional units.

Simultaneous Multithreading (SMT) [1] technology allows a single physical processor core to execute instructions from multiple threads in the same cycle. SMT technology improves the utilization of functional units by increasing ILP and improves the overall system throughput via Thread-Level Parallelism.

However, Single-Thread (ST) performance might suffer because of SMT technology. ST performance degradation presents a significant challenge to using SMT technology with latency-critical applications.

Naturally, thread-level parallelism could be exploited in latency-critical applications and the generated parallel tasks could be scheduled to logical threads of an SMT core. To address challenges of parallelization of general-purpose applications, previous works also explored Thread-Level Speculation on SMT cores [2]. Helper Threading [3] is another technique that could use SMT technology to improve the performance of heavy threads in a latency-critical application.

Wide-spread use of thread-level speculation and helper threading to improve ST performance of heavy threads is challenging. Thread-Level Speculation requires the support for transactional memory, while an efficient helper threading on SMT cores usually requires hardware support.

Previously, in [4], a specialized parallel programming framework, called Relic, was introduced. It enables extremely fine-grained task parallelism on SMT cores.

In this work, we explore methods to accelerate latency-critical applications by parallelizing fine-grained kernels within them using Relic framework. There is a large scope of work on discovery of parallelism in sequential programs. Recently, Large Language Models (LLMs) emerged as the promising approach for identification of potential parallel regions and restructuring of the code. Several works demonstrate that LLM-based solutions could significantly outperform traditional methods [5].

However, previous works showed limitations of general-purpose LLMs, such as GPT models, to produce parallel code and parallelize sequential code. That's why in most of the previous works, specialized LLMs are trained and used for the tasks involving parallel programming and parallelization of programs. Training of these specialized models still require significant computational resources.

Instead of relying on specialized models, we use state-of-

Denis Los – Moscow Institute of Physics and Technology (9 Institutskiy per., Dolgoprudny, Moscow Region, 141700, Russian Federation) ORCID: https://orcid.org/0009-0009-4500-8106 email: los.da@phystech.edu

Igor Petushkov - Moscow Institute of Physics and Technology (9 Institutskiy per., Dolgoprudny, Moscow Region, 141700, Russian Federation) email: piv-tula@mail.ru.

the-art general-purpose models, but take a different approach to improving their efficiency in parallelizing sequential programs. We integrate our solution directly to a popular AI Code Editor, called Cursor [6], and use it with its AI Coding Agent, connecting additional tools to provide additional context to the model

In this paper, we make the following main contributions:
1) We introduce Aira: AI-powered Parallelization Adviser. It's based on the AI Coding Agent used in Cursor IDE with Claude Sonnet 4 model in its core. We develop and connect additional tools to the agent through Model Context Protocol (MCP). Additional functionality that we add include: sample-based profile collection to detect hotspot functions, Dynamic Binary Instrumentation (DBI) tool to collect dynamic information, binary analysis tool to analyze static dependencies and collected execution traces
2) We introduce a specification file describing the end-to-end flow of Aira to the LLM from hotspot detection to parallel restructuring of code. The specification file is loaded into the context of the model through MCP. The optimization flow of Aira could be enabled with a simple prompt message: "Parallelize this program with Aira".
3) We rely on the LLM to annotate promising code regions after hotspot detection and restructure potential parallel code regions with Relic framework at the end. The specification file features examples of the usage of Relic framework to provide the model with the context. This allows to do parallel restructuring of code using a custom parallel programming framework.
4) We develop a binary analysis tool based on Binary Optimization Layout Tool (BOLT) to analyze static dependencies and dynamic dependencies with collected execution traces in the annotated code regions
5) We extend Sniper simulator to analyze a potential benefit from running the tasks on an SMT core based on the collected execution traces
6) With Aira, we automatically parallelize latency-critical applications using Relic framework.

## II. RELATED WORK

There is a wide range of existing parallelization tools coming from decades of research.

LLVM Polly [7] is the classical tool for parallelization of loops and improving affinity integrated into LLVM. It uses an abstract polyhedral model to analyze memory access patterns. The polyhedral model works best for regular memory accesses that represent many cases in scientific computing workloads. GCC has its own framework for polyhedral analysis called Graphite [8].

DiscoPop [9] is another tool of semi-automatic program parallelization. DiscoPoP uses LLVM IR-level instrumentation to assemble a memory access profile from all instructions. The runtime library allows you to monitor memory accesses during execution. In order to reduce delays caused by profile collection, memory accesses are written to a hash table, and instructions with repeated private accesses are skipped.

Apollo [10] tool allows to do automatic, dynamic and speculative parallelization. Apollo features 2 main components. The first component is the set of extensions for Clang compiler. The second component is a runtime system that can do instrumentation to collect information about dynamic memory accesses.

Par4All [11] is an automatic parallelizing and optimizing compiler for C and Fortran programs. It is mostly targeted towards loop parallelization.

PLUTO [12] uses polyhedral model to find coarse-grained parallelism in the big code sections, such as nested loops.

Other tools for automatic and semi-automatic parallelization include Intel Adviser XE, ParaMeter, Prism, SLX Tool Suite.

Recently emerged AI-Driven tools include OMPar [5], that uses two models: OMPify and MonoCoder-OMP. The first one is used to access loop parallelization potential and the second one is used to generate precise OpenMP pragmas. Other AI-driven tools for generating OpenMP pragmas include [13]-[18].

## III. ENVIRONMENT

We conduct all experiments on a Linux system with Ubuntu 24.04. The system has Intel Core i7-12700 Alder Lake processor featuring 8 performance cores and 4 efficient cores. For all experiments on SMT technology, we use performance cores with Hyper-Threading technology. Linux kernel version is 6.14 and glibc is 2.39.

All benchmarks and parallel programming frameworks are compiled with Clang compiler from LLVM 20.1.8 with -O3 optimization option. We use libc++ from LLVM 20.1.8 as the implementation of C++ standard library for all experiments.

We use Cursor IDE v1.3.9 that is based on VS Code 1.99.3.

## IV. ANALYSIS OF THE SMT TECHNOLOGY EFFICIENCY FOR FINE-GRAINED KERNELS

Many previous works show that performance benefit from SMT technology strongly depends on an application [19][20][21]. In general, a pair of tasks running on logical threads of an SMT core should complement each other and avoid competing for the same functional units. Compute-bound kernels fully utilizing available functional units would not see any benefit from SMT technology or would see a negative impact. Memory-bound tasks are more likely to see performance benefit on an SMT core, however, cache contention between the tasks could lead to degradation of performance.

While there have been many works studying the efficiency of SMT technology, they mostly focused on coarse-grained tasks. To understand which tasks should be selected by Aira adviser and study the underlying effects, we analyze sets of compute-bound and memory-bound fine-grained kernels running on SMT cores. The kernels are parallelized with Relic [4] framework and LLVM OpenMP. For both Relic and LLVM OpenMP, only 2 threads are used. The threads are scheduled to either a single physical core with Hyper-Threading or to 2 distinct physical cores. These two scenarios are referred to with suffixes SMT or SMP (which stands for Symmetric Multiprocessing), respectively. We vary granularities of the kernels using the corresponding parameters.

In Fig. 1, for example, performance gains from parallelization with Relic framework and LLVM OpenMP are shown for the motion update in Particle Filter Localization (PFL) benchmark from Real-Time Robotics Benchmark (RTRBench) [22]. This is a compute-bound benchmark with floating-point operations.

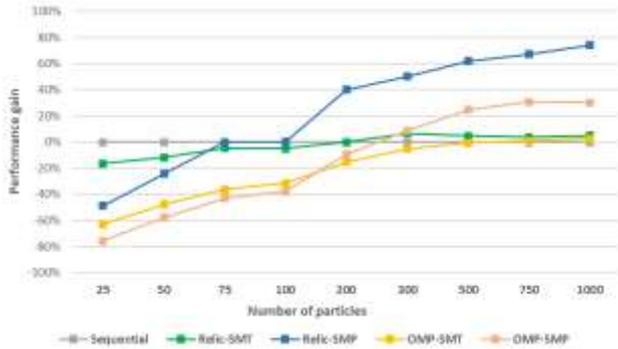

Figure 1. Performance gains from parallelizing motion update in PFL benchmark with different granularities

For motion update in PFL benchmark, there is no significant performance gain from simultaneous multithreading. For 1000 particles, performance gains are 5.1% and 2.7% for parallelization on an SMT core with Relic framework and LLVM OpenMP, respectively.

For small numbers of particles, there is performance degradation from parallelization due to task scheduling overheads. However, since task scheduling overhead is smaller for parallelization on an SMT core, for both Relic and LLVM OpenMP, SMT-based parallelization shows higher performance than parallelization on different physical cores. The performance gains from SMT-based parallelization for small granularities are still negative. As expected, Relic framework shows better performance than LLVM OpenMP on small granularities.

The interesting observation from Fig. 1 is that for a very short range of granularities, SMT-based parallelization with Relic framework shows higher positive performance gain than SMP-based parallelization with LLVM OpenMP.

In Fig. 2, performance gains from parallelization with Relic framework and OpenMP are shown for the CC benchmark introduced in [4]. It's a fine-grained memory-bound graph processing benchmark.

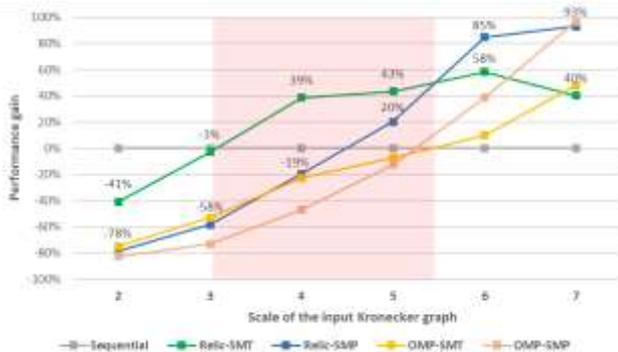

Figure 2. Performance gains from parallelizing CC benchmark with different granularities

For very small granularities, there is still performance degradation for parallelization due to not low enough task scheduling overheads for the memory-bound CC benchmark. However, there is a range of granularities, for which performance speedup from parallelization was made possible by the introduction of Relic framework. This is the range where SMT-based parallelization with Relic framework shows positive performance gains and outperforms SMP-based parallelization. In the same range of granularities, both SMT-based and SMP-based versions of parallelization with LLVM OpenMP result in performance drop due to high task scheduling overheads. Below this range of granularities, all versions of parallelization result in performance degradation. Above this range of granularities, it's more beneficial to schedule parallel tasks to different physical cores. These range of granularities represent the primary target of fine-grained parallelization on SMT cores.

Hence, as expected, to effectively parallelize fine-grained kernels of latency-critical application, Aira adviser should be able to characterize the kernels and estimate their granularity.

## V. AIRA: AI-POWERED PARALLELIZATION ADVISER

Aira is based on the AI Coding Agent used in Cursor IDE. We choose Claude Sonnet 4 model to power the agent, since its outstanding performance on coding benchmarks.

We develop a specification in Markdown document for the end-to-end flow of applying Aira. A special MCP tool loads this document into the context of the model when a user inputs a prompt asking for parallelization with Aira.

The specification file starts from collecting sampled profiles with Linux perf tool with enabled LBR feature. We develop a wrapper tool to collect sampled profiles, parse them and provide to the LLM in JSON format representing the found hot functions.

Then the model is instructed in the specification file to annotate promising for parallelization code regions inside of hotspot functions. The mapping of each annotated region to the source code line positions is saved in the additional file.

After adding instrumentation, execution traces featuring load and store memory accesses with executed basic blocks are collected with DynamoRIO.

A program with annotations is passed to a binary analysis tool. We developed this tool based on Binary Optimization Layout Tool (BOLT), that is used in LLVM. BOLT is mainly used for code layout optimization; however, it provides an extensive framework for binary analysis and optimization going beyond optimization for code layout.

For each annotated code region, we check static dependencies using BOLT, as well as dynamic dependencies (memory accesses) using collected execution traces. The binary analysis tool can work with binaries optimized with -O3 option since it only looks for the annotation marks. However, the binary analysis tool could also report found memory dependencies mapping them to the original variables in the source code through the llvm-symbolizer tool.

If there are no conflicts detected during the binary analysis, potential performance benefit is estimate based on trace-based Sniper simulator. Sniper simulator, based on Sniper 7.4 was extended to more accurately model allocation of issue ports and several functional fixes were applied to the SMT model of Sniper. A simplified OOO model is used in Sniper instead of the interval model for more accuracy. If there are no conflicts for the code region and potential performance benefit is reported with the Sniper performance simulator, the output is generated for the LLM.

The next command in the specification files asks for parallel restructuring of the found regions with Relic parallel programming framework.

Since the LLM is not aware of the interface of Relic framework, we include 20 simple examples of applying Relic framework to various parallelization cases. It shows possibility to apply Aira to custom frameworks and go beyond OpenMP parallel programming model.

## VI. BENCHMARKS

We have chosen 10 benchmarks that represent a wide range of real-world latency-critical applications from different domains, such as cybersecurity, high-frequency trading (HFT), robotics, social media, recommendation systems, and aerospace. All of these benchmarks use algorithms based on linked data structures and may show high cache miss rates.

### A. Geo-Spatial Database System (GeoSpatial)

This benchmark implements a latency-critical geo-spatial key-value query engine that simulates the critical path of a geo-spatial database system. Three sequential stages comprise the query processing system. First, spatial range queries are performed using a k-dimensional (KD-tree) data structure to identify objects within rectangular regions through iterative tree traversal. Second, the metadata associated with spatially identified objects is retrieved using a binary search tree. Third, for each query, an aggregated value is computed after processing linked lists with per-object metrics.

In the database, we keep 2048 geo-spatial objects, distributed across a 1000 x 1000 coordinate space. We perform 1000 warmup iterations and $10^5$ measurement iterations, where each iteration processes 15 concurrent rectangular range queries. Rectangular queries span 50 x 50 coordinate units. We limit the maximum number of identified objects per query to 32.

### B. Volume-Weighted Average Price Engine (VWAP)

The volume-weighted average price engine benchmark implements a latency-critical Volume-Weighted Average Price (VWAP) computation engine designed for high-frequency trading market data analytics. The processing pipeline looks like the following. First, the incoming trade prices are mapped to discrete order book levels. Skip-list data structure is used to perform these search operations efficiently. Then, in the volume aggregation stage, linked-list traversal is performed to aggregate outstanding volume at each identified price level. Finally, volume-weighted average prices are computed over a sliding window using circular ring buffer traversal.

Price levels show uniform distribution across $100.00-$100.99 range with a 1-cent minimum price increment. The skip-list has four hierarchical levels. The sliding window size is set to 32 ticks, representing approximately 32-64ms of market activity at typical message rates. We perform 1000 warmup iterations and $10^5$ measurement iterations, where 30 concurrent trade messages are processed on each iteration.

### C. Obstacle Detection System (LIDAR)

This benchmark implements a real-time obstacle detection and collision system designed for autonomous vehicle safety applications. 3D point cloud data from LIDAR sensors is processed to determine minimum obstacle distances along planned vehicle trajectories. The system constructs a balanced KD-tree data structure from the point cloud observations to enable efficient processing of spatial queries. To resolve a safe trajectory, nearest-neighboring queries are performed.

Point cloud observations are distributed within a 60 x 60 x 60 meter cubic sensing volume, representing a typical scenario in an urban area. We use 1000 obstacles that are uniformly distributed throughout the sensing volume, simulating pedestrians, vehicles, and static infrastructure. The trajectory is represented using 100 discrete waypoints with 0.2-meter spatial resolution, resulting in a forward-looking trajectory of up to 20 meters. We use 1000 warmup iterations and $10^5$ measured iterations.

### D. Social Media Feed Generation System (Timeline)

The Timeline benchmark implements a high-performance social media feed generation microservice designed to emulate production-scale content recommendation systems deployed by major platforms such as Twitter and LinkedIn. Potential content posts are collected from followed accounts after traversing the viewer's social graph. The collected potential content posts are evaluated based on engagement metrics and temporal decay functions.

For the Timeline benchmark, we use 1000 active accounts in the social graph. Each user follows from 64 to 192 other accounts and maintains from 16 to 80 posts in their timeline. Each post could receive from 5 to 25 reactions. The number of posts and followed accounts for each user, as well as the number of reactions for each post, is determined based on a uniform random distribution. Post timestamps are distributed across the previous 24-hour period to model temporal content distribution. The maximum number of content posts from each user is limited to 8.

### E. Random Forest (RF)

This benchmark implements a Random Forest ensemble consisting of multiple binary decision trees, each represented as a linked data structure with internal nodes containing feature indices, thresholds, and pointers to child nodes.

The number of decision trees in the ensemble is equal to 256. Each decision tree has a maximum depth of 5. Each input feature vector has 32 features. We use 1000 warmup iterations and $10^5$ measurement iterations.

### F. Graph Neural Network 1-Hop Embedding (1-Hop)

For this benchmark, the focus is on evaluating the performance of graph neural network inference systems. The computation of 1-Hop embeddings is simulated in this benchmark, a core operation in applications such as social network analysis and recommendation systems.

The graph used in this benchmark has 200000 nodes with an average degree of 256. There are 64 features per node. For consistency, on each iteration, the embeddings are computed for a node with index 0. The number of iterations is $10^5$.

### G. Limit Order Book (LOB)

The LOB benchmark evaluates the performance of high-frequency trading (HFT) matching engines that process limit order book updates in real-time financial markets. This benchmark simulates a multi-symbol order book management system where each symbol maintains its own price-level structure, representing the core computational kernel of latency-critical trading infrastructure. The benchmark implements a limit order book (LOB) data structure that maintains price levels in ascending order, with each level containing a linked list of orders.

For the LOB benchmark, we use 256 independent trading symbols. We set the number of order updates per symbol per iteration to 500. The total number of updates is equal to 128000. Prices are distributed uniformly in the range from 100\$ to 101\$. We use 100 warmup iterations and 1000 measurement iterations.

### H. IP Address Geolocation System (GeoIP)

The GeoIP benchmark implements an ultra-low-latency IP address geolocation system commonly deployed in content delivery networks, firewalls, and edge computing infrastructure. This benchmark features a binary trie data structure optimized for IPv4 address lookup, where each node represents a bit position in the 32-bit IP address.

On each iteration, we process $10^6$ IP addresses. We use 100 warmup iterations and 1000 measurement iterations.

### I. Fraud Detection (Fraud)

This benchmark evaluates the performance of graph-based anomaly detection systems that identify complex network motifs in large-scale transaction or interaction graphs. This benchmark implements a 5-vertex fan-in motif detection algorithm that identifies suspicious patterns where multiple entities converge on a single target, representing the computational kernel of real-time fraud detection and network security systems.

For this benchmark, we use a graph with $10^5$ vertices and $3 \cdot 10^5$ random background edges. We use 100 warmup iterations and 1000 measurement iterations. In each iteration, 1000 edges are tested for pattern completion.

### J. 3D Collision Detection System (BVH)

The BVH benchmark evaluates the performance of 3D collision detection systems used in autonomous vehicles, robotics, and aerospace applications. This benchmark implements a Bounding Volume Hierarchy (BVH)-based collision detection algorithm that identifies potential collisions between a predicted trajectory and a point cloud of obstacles, representing the computational kernel of real-time safety systems.

For the BVH benchmark, we use $2 \cdot 10^5$ obstacle points in 3D space. We evaluate 10000 trajectory points. For the environment, we use a 1 km cube. Obstacles are distributed uniformly across 3D space. We use 100 warmup iterations and 1000 measurement iterations.

## VII. RESULTS

We apply an end-to-end parallelization pipeline of Aira to optimize latency-critical benchmark applications. 7 out of 10 latency-critical benchmarks were successfully automatically parallelized with Relic framework after using Aira. In Fig. 3, performance gains are shown for each benchmark with the positive performance impact. Geomean performance gain on the benchmarks with positive performance impact is 25.2%.

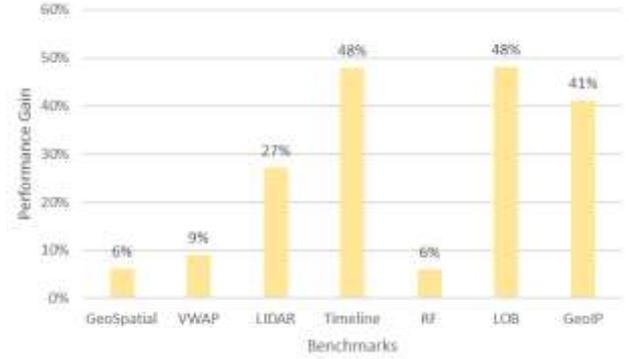

Figure 3. Performance gain on latency-critical benchmarks after applying Aira with Relic framework

3 out of 10 benchmarks could not be successfully parallelized using Aira with Relic framework. Performance degradation results for each benchmark are shown in Fig. 4. Parallelization of Fraud detection benchmark was not found to be beneficial during performance simulation of the SMT Core in Sniper simulator. Hence, Relic framework was not applied and there is no change in the performance of the Fraud benchmark. However, for 1-Hop and BVH benchmarks were not flagged during the check in Sniper simulator, however, the parallelized kernels were too fine-grained to apply Relic framework. Performance degradation in the BVH benchmark is 61%. Negative performance impact in the 1-Hop benchmark is 9%. Real-world latency-critical application are extensively profiled; hence, performance degradations are usually discovered and features with negative impacts are discarded. Without negative outliers, the geomean performance gain from applying Aira with Relic framework is 17%.

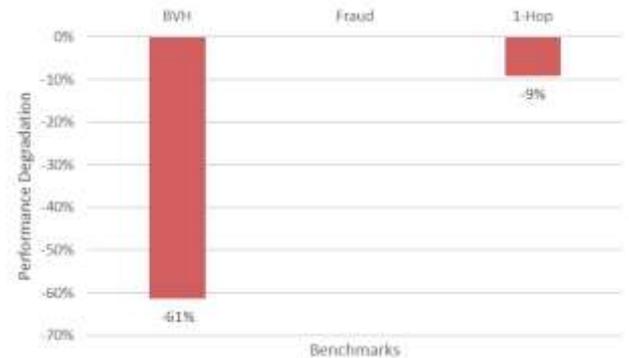

Figure 4. Performance degradation on some of the latency-critical benchmarks after applying Aira with Relic framework

## VIII. Conclusion

We introduced Aira, an AI-powered parallelization adviser. Instead of relying on the specialized LLMs, we used as the foundation AI Coding Agent used in Cursor IDE that is powered by Claude Sonnet 4 model. We developed and integrated additional tools via MCP to do hotspot detection, collect and analyze dynamic dependencies. We introduced a specification Markdown file describing the whole optimization flow of Aira so that Aira could be easily used without the need to input complex prompts.

We analyzed efficiency of SMT technology for fine-grained kernels and extended Sniper performance simulator to estimate performance gain from the parallelization on SMT cores based on the collected execution traces and discard parallelization cases with negative performance impact.

We applied Aira with Relic framework to 10 latency-critical benchmarks representing real-world industry applications and use cases and achieved average performance gain of 17%.